\documentclass[fleqn,usenatbib]{mnras}

\usepackage{newtxtext,newtxmath}

\usepackage[T1]{fontenc}

\DeclareRobustCommand{\VAN}[3]{#2}
\let\VANthebibliography\thebibliography
\def\thebibliography{\DeclareRobustCommand{\VAN}[3]{##3}\VANthebibliography}

\usepackage{graphicx}	
\usepackage{amsmath}	

\title[Distance and metalicity of Bulge clusters]{Distance and [Fe/H] of Galactic bulge clusters from member RR Lyrae $I$-band light curves}

\author[Arellano Ferro \& Prudil]{
A. Arellano Ferro$^{1}$\thanks{E-mail: armando@astro.unam.mx}, Z. Prudil$^{2}$ \thanks{Equal first authors.}
\\
$^{1}$Instituto de Astronom\'ia, Universidad Nacional Aut\'onoma de M\'exico, Ciudad de M\'exico, CP 04510, M\'exico.\\
$^{2}${European Southern Observatory, Karl-Schwarzschild-Straße 2, 85748, Garching, Germany}\\
}

\date{Accepted -- 27. Received --; in original form --}

\pubyear{2020}

\begin{document}
\label{firstpage}
\pagerange{\pageref{firstpage}--\pageref{lastpage}}
\maketitle


\begin{abstract}
We have investigated the results for [Fe/H] and distance for a group of 24 globular clusters in the Galactic bulge, employing recent calibrations of RR Lyrae light curves Fourier decompositions and period- absolute magnitude -metallicity (PMZ) calibrations in the $I$-band. We have limited our calculations to RR Lyrae stars that have been proven to be very likely cluster members. These results are compared with [Fe/H] and $M_V$ (distance) obtained from well-established Fourier calibrations in the $V$-band. These calibrations of the $I$-band were found to produce iron values that can differ from the UVES spectroscopic scale by -0.29 to +0.15 dex. The PMZ distances agree within 0.4 kpc with recent solid critical distance compilations. Adopting the newly derived distances, we conducted a spatial and orbital analysis of the bulge globular clusters in a non-axisymmetric Milky Way potential, and compared their orbital properties with earlier studies, finding broadly consistent trends with small systematic differences driven by the assumed distances and Galactic model. Clusters associated with the in situ bulge component
display a narrow range low angular
momentum and low orbital energies, consistent with
formation in the early inner Milky Way.
\end{abstract}



\section{Introduction}

The road towards the determination of physical parameters of fundamental relevance in pulsating 
stars from the morphology of their light curves has been steadily broadening over the last sixty years.

In his preliminary study of non-linear hydrodynamical models of stellar envelopes of RR Lyrae stars, \citet{Christy1963} made clear that
{\it "... the
calculated envelopes of appropriately chosen models have the appearance 
of actual RR Lyrae stars, including the correct shapes of
light and velocity curves"}.
The first hydrodynamical models aimed to study the envelopes of
RR Lyrae stars were built by \citet{Christy1966} whom explored the effects of 
varying the mass, mean radius, luminosity and hydrogen-to-helium
ratio, revealing details of the shapes of light- and velocity-curves.
Further work aimed at inferring the stellar mass and luminosity as function of composition, 
assumed for the models of Horizontal-Branch stars, from theoretical and observational grounds, 
was
developed by \citet{vanALBADA1971}; \citet{vanHerk1971}. It has been clear since the above seminal works, that the light curve morphology in RR Lyrae stars constitutes a wealth of information 
on the fundamental physical parameters, particularly the metallicity ([Fe/H]) and luminosity.
Early initiatives to matching the light curve morphology, via their Fourier decomposition, with 
hydrodynamic pulsation models, lead \citet{Simon1988} and \citet{SimonClement1993}
to identify proper numerical calibrations to derive masses, luminosities and temperatures in RRc stars.

 A few years later, empirical calibrations of accurate observables (period and the Fourier parameter of $V$-band light curves) and individual stellar parameters were calculated for the iron abundance [Fe/H] \citep{Jurcsik1996} (JK96) and the absolute magnitude $M_V$ \citep{Kovacs2001} of fundamental mode RR Lyrae. For the first overtone RR Lyrae, 106 stars in globular clusters yielded a very useful calibration for the iron abundance in the Zinn-West scale \citep{Morgan2007}. For the above iron abundance calibrations, abundances are based mostly on the $\Delta S$ method of \citet{Preston1959}.

The effective temperature $T_{\rm eff}$ can also by calculated from Fourier parameters from the calibrations of \citet{SimonClement1993} for RRc stars and \citet{Jurcsik1998} for RRab stars. Some caveats regarding the RRc calibration are discussed by \citet{Arellano2010}, who also calculates the Fourier-based individual stellar radii and masses.

The specific equations of the above calibrations have been listed by \citet{Arellano2010} where special attention was paid to the zero points of the absolute magnitude calibrations. The homogeneous iron values and distances for a family of 39 globular clusters are reported by \citet{Arellano2024} where it is shown that the matallicities and distances from the Fourier decomposition of the $V$-band light curves, and the above calibrations are consistent with the spectroscopic scale of \citet{Carretta2009}, [Fe/H]$_{\rm UVES}$ and the distance scale of \citet{Baumgardt2021}, $d_{BV}$.

Based on values of [Fe/H] obtained from high resolution spectroscopy and accurate $I$-band photometry for 28 stars, \citet{Smolec2005} produced a calibration that, like the $V$-band calibration of JK96, expresses [Fe/H] in terms of the period and the $\phi_{31}$ parameter. This calibration could be of great value considering the abundant and accurate $I$-band light curves of RR Lyrae stars in the extensive photometric survey of the Galactic bulge, Optical Gravitational Lensing Experiment (OGLE) data base \citep{Udalski1992,Udalski2015},
In addition,  \citet{Smolec2005} offered a three-parameter calibration involving the amplitude of the second harmonic $A_2$ (see his equations 2 and 3). Later, \citet{Hajdu2018} argued that the three-parameter formulation yields more consistent abundance estimates.

A small modification to the [Fe/H] calibration of \citet{Smolec2005} has been proposed by \citet{Jurcsik2021} to improve the consistency of photometric metallicities with the spectroscopic UVES scale of \citet{Carretta2009} for RRab stars. An independent calibration of the Fourier parameters in the $I$ band was calculated by \citet{Dekany2021}, based on 96 RRab and 34 RRc calibrators with high-resolution spectroscopy collected from the literature and well-covered $I$ light curves taken from \citet{Monson2017} and All Sky Automated Survey (ASAS) \citep{Pojmanski1997}. OGLE $I$-band light curves were not used for calibration purposes since very few (2) stars were in common with the spectroscopic sample.

Period-absolute magnitude-metallicity (PMZ) relations in assorted passbands have been calculated by \citet{Prudil2024B} for more than $100$ calibrators with iron abundances from the literature and parallaxes determined by the \textit{Gaia} satellite. These relations and the reddening map newly created for the Galactic bulge by \citet{Prudil2025}, allowed these authors to calculate iron abundances and distances for more than seventy thousand stars in the Galactic bulge. Recent \textit{Gaia}-based analyses have established astrometrically confirmed samples of RR~Lyrae and type-II Cepheid members in Galactic globular clusters, enabling homogeneous high-precision distance calibrations via Leavitt laws. These studies also provide robust probabilistic frameworks for the selection of the RR~Lyrae cluster membership and highlight current limitations of the photometric metallicity estimates \citep{Cruz2024,Lengen2025}.

In the present paper, we aim to determine, in a homogeneous approach, the iron abundance and distance to bulge globular clusters, using the above-mentioned $I$-band iron calibrations and the PMZ relations, applied exclusively to critically select likely cluster members RR Lyrae stars. In section \S 3 the Fourier light curve decomposition is described, and the specific calibrations towards the calculation of [Fe/H] and distance are presented. Sections \S 4 and 5 contain some membership considerations and the inclusion criteria for stellar membership. In \S 6 the resulting iron abundances and distances are presented, and the different scales involved are compared and evaluated for consistency. In section \S 7 our newly derived distances and a non-axysimmetric Galactic potential are used to calculate the spatial and kinematical distribution of our sample of bulge clusters.
Section \S 8 summarises our conclusions. Finally, in Appendix A we highlight some pecular stars in several clusters.

\section{RR Lyrae $I$-band light curves Fourier decomposition: Specific equations for [Fe/H] and distance}
\label{sec:Four}

We employed the $I$-band light curves available in the OGLE IV database for RRab and RRc stars present in a family of Galactic bulge globular clusters. Stars with obvious Blazhko amplitude modulations were avoided. Although generally the OGLE's $I$ light curves are of excellent quality and are fully covered in phase, for some faint clusters, stars near the cluster central regions are blended and exhibit large scatter. These were also not included in the average metallicity and distance calculations.

The $I$-band light curves were fitted with a Fourier series harmonics model of the form:

\begin{equation}
\label{eq.Foufit}
I(t) = I_0 + \sum_{k=1}^{N}{A_k \cos\ ({2\pi \over P}~k~(t-E) + \phi_k) },
\end{equation}

\noindent
where $I(t)$ is the $I$-band magnitude at time $t$, $P$ is the period, and $E$ is the epoch. A
linear
minimization routine is used to derive the best-fit values of the 
amplitudes $A_k$ and phases $\phi_k$ of the sinusoidal components. 
From the amplitudes and phases of the harmonics in eq.~\ref{eq.Foufit}, the 
Fourier parameters, defined as $\phi_{ij} = j\phi_{i} - i\phi_{j}$, and $R_{ij} =
A_{i}/A_{j}$, are computed. The Fourier coefficients are not specifically listed in this paper, but they are available on request.

We employed two independent calibrations to estimate [Fe/H] from the low-order Fourier parameters from the $I$ light curves decomposition; the calibration for RRab stars of \citet{Smolec2005} slightly modified
by \citet{Jurcsik2021} (their eq. 3):

\begin{equation}
\label{eq:J21}
{\rm[Fe/H]}J_{\rm ab} = -6.018 -4.261 P+ 1.120 \phi_{31} + 7.466 A_2,
\end{equation}

\noindent
and the calibrations for RRab and RRc stars of \citet{Dekany2021} (their table 6):

\begin{equation}
\label{eq:D1}
{\rm[Fe/H]}D_{\rm ab} = -5.819 -6.350 P + 1.248 \phi_{31} + 5.785 A_2,
\end{equation}

\noindent
and

\begin{equation}
\label{eq:D2}
{\rm[Fe/H]}D_{\rm c} = -1.821 -10.014 P + 0.325 \phi_{31} -34.704 A_2 +13.835 A_1.
\end{equation}

For the calculation of the distance, we made use of the PMZ calibration to estimate the absolute magnitude in the $I$-band, $M_I$ derived by \citet{Prudil2024B} (their equation 19):

\begin{equation}
\label{eq:P24}
M_I = -1.292 log_{10} P +0.196 {\rm [Fe/H]} +0.197.
\end{equation}

With the absolute magnitude and an assumption of the interstellar extinction $R_{VI}$, the distance in parsecs emerges as:

\begin{equation}
\label{eq:Dist}
d_I=10^{(1+ 0.2 (I_0-M_I-R_{VI}*E(V-I))}.
\end{equation}

\noindent
The value $R_{VI}=1.205$ was adopted \citep{Prudil2025}. The colour excess $E(V-I)$ is rather more controversial. It is a well-known fact that bulge clusters are not only subject to heavy interstellar extinction but also that extinction is of a differential nature.

A detailed analysis of the extinction variations in the Galactic bulge has recently been offered by \citet{Prudil2025}. 
Reddening maps were generated, and new extinction laws were derived from the visual to near infrared bands, for many of the more than 70 thousand individual stars considered in their study. It was found that the extinction law is more uniform in $R_{IK}$ and $R_{JK}$. The colour excesses  $E(I-K)$ and $E(J-K)$ are provided for stars with available near infrared photometry from the Vista Variables in the Via Lactea survey \citep{Minniti2010} or the Vista Hemisphere Survey \citep{McMahon2013}.
These results cover a tremendously large number of RR Lyrae stars in the Galactic bulge, for which individual distances were calculated. Many of such RR Lyrae are in the field of globular clusters, but they are not necessarily cluster members (see discussion in \S4).

During our detailed exploration of the RR Lyrae stars associated to globular clusters, we encounter that some of the known RR Lyrae in
globular clusters according to the Catalogue of Variable Stars in Globular Clusters (CVSGC) \citep{Clement2001}, do not have reddening values in \citet{Prudil2025} and hence no distance estimations, most likely due to lack of $JK$ photometry. This circumstance prevented us from a star-by-star reddening and distance estimation and opted instead for adopting an average reddening for each cluster in our sample, from \citet{HARRIS1996}. The resulting mean metallicities and distances for the Bulge clusters in our sample, following the above calibrations, are given in Table \ref{Results} and are discussed in \S 5.

\section{Stellar membership Considerations}
\label{sec:membership}

RR Lyrae stars can be used as indicators of average metallicity and distance of a given globular cluster if it can be argued that they are physically associated. Establishing cluster membership is a difficult task that has been much alleviated since the paramount collection of proper motions and parallaxes measured by the $Gaia$-DR3 mission \citep{Gaia2023}.

For each cluster in our sample, we have limited our calculations to those RR Lyrae that, from their proper motion analysis and their mean brightness, are most likely cluster members. The proper motions analysis from \citet{Vasiliev2021} is the support of the membership assessment of each known variable in Galactic globular clusters performed by \citet{Prudil2024}, now coded in the CVSGC \citep{Clement2001}. We have then used such codification as the membership guide of RR Lyrae stars in the clusters in our sample. It is worth noting at this point that the number of actual field variables assigned to bulge clusters in the CVSGC is quite large, a fact that should not be neglected.

\begin{table*}
\centering
    \caption{Metallicity and distance of bulge clusters from $I$-band light curve Fourier decomposition}
\label{Tab:DistanceToClusters}
\begin{tabular}{ccccccccc}
\hline
Cluster & [Fe/H]$_J$  & [Fe/H]$_D$ &$d_I$ &Nab & Nc & $d_P$ &[Fe/H]$_{\rm UVES}$ & d$_{BV}$ \\
NGC &  & &kpc& & &kpc && kpc   \\
\hline
6266&-0.970$\pm$0.064 &-1.368$\pm$0.111  &6.83$\pm$0.52  & 20 & 9  &6.19$\pm$0.31 (29) &-1.18$\pm$0.07 &6.03$\pm$0.09\\

6284&-1.001   &-1.502   &13.98   & 1& 0  &14.12 (1)  &-1.31$\pm$0.09  &14.21$\pm$0.42\\

6287&{\it -1.812}   &-2.419  &9.84   &{\it 1}& 1  &7.95  (1)& -2.12$\pm$0.09  &7.93$\pm$0.37 \\

6293&--   &-2.12   &9.27&0 & 1  & 9.74 (1)&-2.01$\pm$0.14   &9.19$\pm$0.28 \\

6304&NO VARS   & -- & -- &--  & --  & -- &-- & --\\

6333&-1.399$\pm$0.063 &-1.949$\pm$0.128  &8.63$\pm$0.49  & 2 & 2  &8.64$\pm$0.16 (4) &-1.79$\pm$0.09 &8.30$\pm$0.14\\

6355&-1.182$\pm$0.263    &-1.832$\pm$0.220   &8.73$\pm$0.14   &3 &0   &8.48$\pm$0.27 (3)  &-1.33$\pm$0.14   &8.66$\pm$0.22 \\

6401&-1.016$\pm$0.064 &-1.369$\pm$0.063 & 8.50$\pm$0.66 &11  & 3  &7.71$\pm$0.20 (14)&-1.01$\pm$0.14 &7.44$\pm$0.22 \\

6441*&--  & --&-- &3 &1  &14.85$\pm$0.61 (4) &-0.44$\pm$0.07 &12.73$\pm$0.16\\

6453&-1.379$\pm$0.059   &--2.056$\pm$0.069   &10.89$\pm$0.82   &2 & 0  &10.74$\pm$0.08 (2) & -1.48$\pm$0.14  &10.07$\pm$0.22 \\

6522& {\it -1.012} &-1.360$\pm$0.021 &8.23$\pm$0.12  & {\it 1} &  3 &8.85$\pm$0.24(3)&-1.34$\pm$0.08  &7.30$\pm$0.21\\

6540&{\it -1.193}  &{\it -1.408} &{\it 5.83}& {\it 1}&0   &3.17 (1)  &-1.02+  &5.91$\pm$0.27 \\

6544&-1.415   & -1.952  &2.92   &1 &0   &2.59 (1)  & -1.47$\pm$0.07  &2.58$\pm$0.60 \\

6558&-0.903$\pm$0.175  & -1.353$\pm$0.216&8.07$\pm$0.17 &3 &1 & 8.91$\pm$0.10 (4)&-1.37$\pm$0.14 &7.79$\pm$0.18\\

6569& -0.918$\pm$0.006 &-1.245$\pm$0.082 &11.11$\pm$0.34 & 2 & 6 &11.36$\pm$0.23 (8) &-0.72$\pm$0.14 &10.53$\pm$0.26\\

6626& -1.104$\pm$0.097 &-1.538$\pm$0.193&5.67$\pm$0.11 &  4 & 1 &5.59$\pm$0.06 (4)&-1.46$\pm$0.09 & 5.37$\pm$0.10 \\

6638& -0.924$\pm$0.130  & -1.258$\pm$0.187  &9.59$\pm$0.23  &2 &5   &10.06$\pm$0.48 (7) & -0.99$\pm$0.07  &9.78$\pm$0.34 \\

6642& -0.961$\pm$0.110 &-1.321$\pm$0.076&8.04$\pm$0.15 &  4 & 4 &7.90$\pm$0.43 (8) &-1.19$\pm$0.14 & 8.05$\pm$0.20 \\

6656&-1.471$\pm$0.072   &-2.051$\pm$0.361  & 3.36$\pm$0.13 & 4 & 7  &3.39$\pm$0.09 (11)  &-1.70$\pm$0.08  &3.30$\pm$0.40  \\

6681&-1.471  &-1.785$\pm$0.312   & 9.10$\pm$0.34  &1 & 1  &8.76$\pm$0.44 (2) &-1.62$\pm$0.08   &9.36$\pm$0.11 \\

6715& -1.387$\pm$0.187 &-1.786$\pm$0.318&26.58$\pm$1.30 & 61 & 11 & 27.74$\pm$2.84 (42) & -1.44$\pm$0.07 & 26.28$\pm$0.33 \\

Ter 1& -0.831$\pm$0.234 &-1.291$\pm$0.415&8.81$\pm$0.17&  2& 0 &6.91$\pm$ 0.50 (2) &-1.29$\pm$0.09 & 5.67$\pm$0.17 \\

Ter 10& -1.438$\pm$0.107 &-2.094$\pm$0.101&8.87$\pm$1.40& 4 & 0 &10.59$\pm$0.31 (3) &-1.62+ & 10.21$\pm$0.40 \\

Djorg 2& -0.939 &-1.297$\pm$0.355   &8.34$\pm$0.11  &1 & 1  &8.22$\pm$1.0 (2) & -1.07+ &8.76$\pm$0.18 \\
\hline

\end{tabular}

\label{Results}
\center{Notes: [Fe/H]$_J$ from calibration by \citet{Jurcsik2021} (eq. \ref{eq:J21}); [Fe/H]$_D $ from calibration by \citet{Dekany2021} (eqs. \ref{eq:D1} and \ref{eq:D2}). Numbers in italic indicate values from a single and likely blended  star.  \\}
            \center{* Oo III cluster, calibrations are likely not valid}
   
   \center{+ Value from \citet{Schiavon2024}}   
   
\end{table*}

\section{Stellar inclusion criteria}
\label{inclusion}

Each globular cluster in our sample contains an assorted number of variable stars in the CVSGC. A cross-match with the OGLE catalog revealed those stars with $I$-band light curve, which we then used to perform a Fourier harmonics decomposition. However, not all the variable stars listed in the CVSGC are cluster members; their light curves may display substantial scatter or Blazhko-like modulations and may be substantially blended, particularly in the crowded regions. In order to include the star in the estimation of mean metallicity and distance for each cluster, we imposed four criteria; 1) the star is a cluster member according to the Criteria of \citet{Prudil2024}, coded in the CVSGC, 2) the star is near the Horizontal Branch, 3) displays a low scatter and has no obvious Blazhko modulations and 4) the blending parameter ipd-fract-multi-peak \citep{Prudil2025} $< 10$.

The blending parameter ipd-fract-multi-peak is a percentage indicator of the number of cases in which a given source had multiple peaks in the PSF. Thus, the criterion ipd-frac-multi-peak $> 10$ removes stars that, in more than 10\% of the cases, showed more than a single PSF peak in processing. The inclusion of this fourth criterion removes a number of otherwise potential stars for the average but reduces the individual RMS. The price we pay in exchange is that we lose a few clusters with one but blended RRL star (NGC 6287, NGC 6522, and NGC 6540). For the sake of completeness, we included these stars in the calculation but highlighted their values in Table 1 with numbers in italic font.

\section{Distance and metallicity results}
\label{sec:Results}

In Table \ref{Results} we list our resulting [Fe/H] and distance estimates from the $I$-band light curve Fourier decompositions calculated as described in \S 3. The detailed content of the table is the following;
Column 2 displays [Fe/H]$_J$ calculated with equation \ref{eq:J21} for the cluster member RRab stars. Column 3 shows the average [Fe/H]$_D$ calculated with equations \ref{eq:D1} or \ref{eq:D2} for  RRab or RRc stars, respectively. Column 4 carries the distance calculated with equations \ref{eq:P24} and \ref{eq:Dist}. Columns 5 and 6 list the number of  RRab and RRc stars involved in the calculations of previous columns. Column 7 is the average of distance given by \citet{Prudil2025} for the stars satisfying the four criteria defined above (number of stars in parentheses). Column 8 gives the spectroscopic iron values of \citet{Carretta2009} and column 9 lists the distance of \citet{Baumgardt2021}. These last two columns are included since it is of interest to compare the values for [Fe/H] and distance, obtained from photometric arguments, with solid results obtained from different approaches. For the case of the iron abundance, the UVES spectroscopic scale of \citet{Carretta2009} is now a classic and respected reference, while the critical estimation of accurate distances of globular clusters by \citet{Baumgardt2021} stands as good bench mark.

\begin{figure*}
\begin{center}
\includegraphics[width=10.0cm]{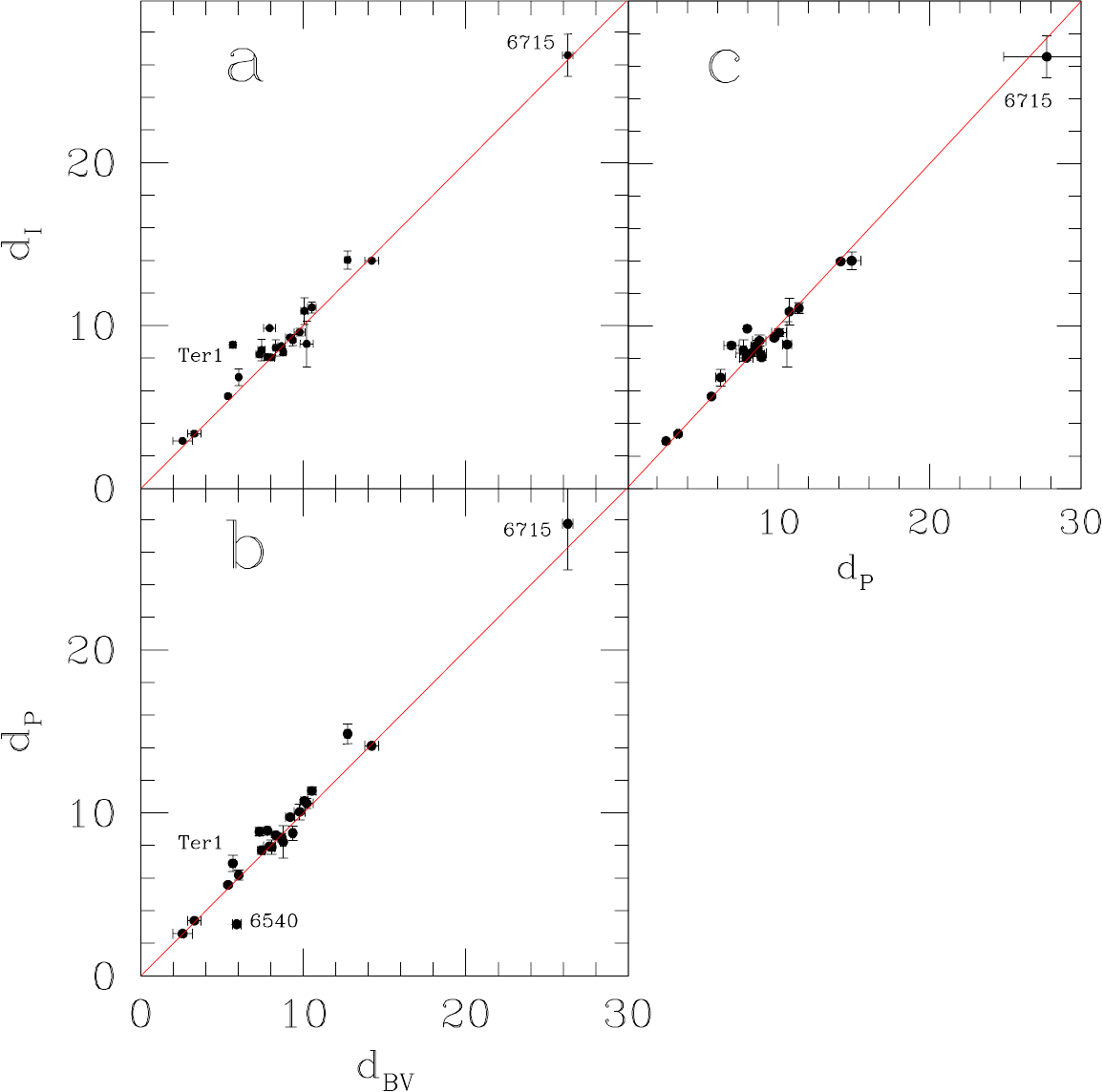}
\caption{Comparison of distances. Panel $a)$ $d_I$ vs $d_{UV}$, panel $b)$ $d_P$ vs $d_{UV}$ and panel $c)$ $d_I$ vs $d_P$. See \S \ref{sec:Results} for a discussion}
\label{Comp_Distance}
\end{center}
\end{figure*}

\begin{figure*}
\begin{center}
\includegraphics[width=10.0cm]{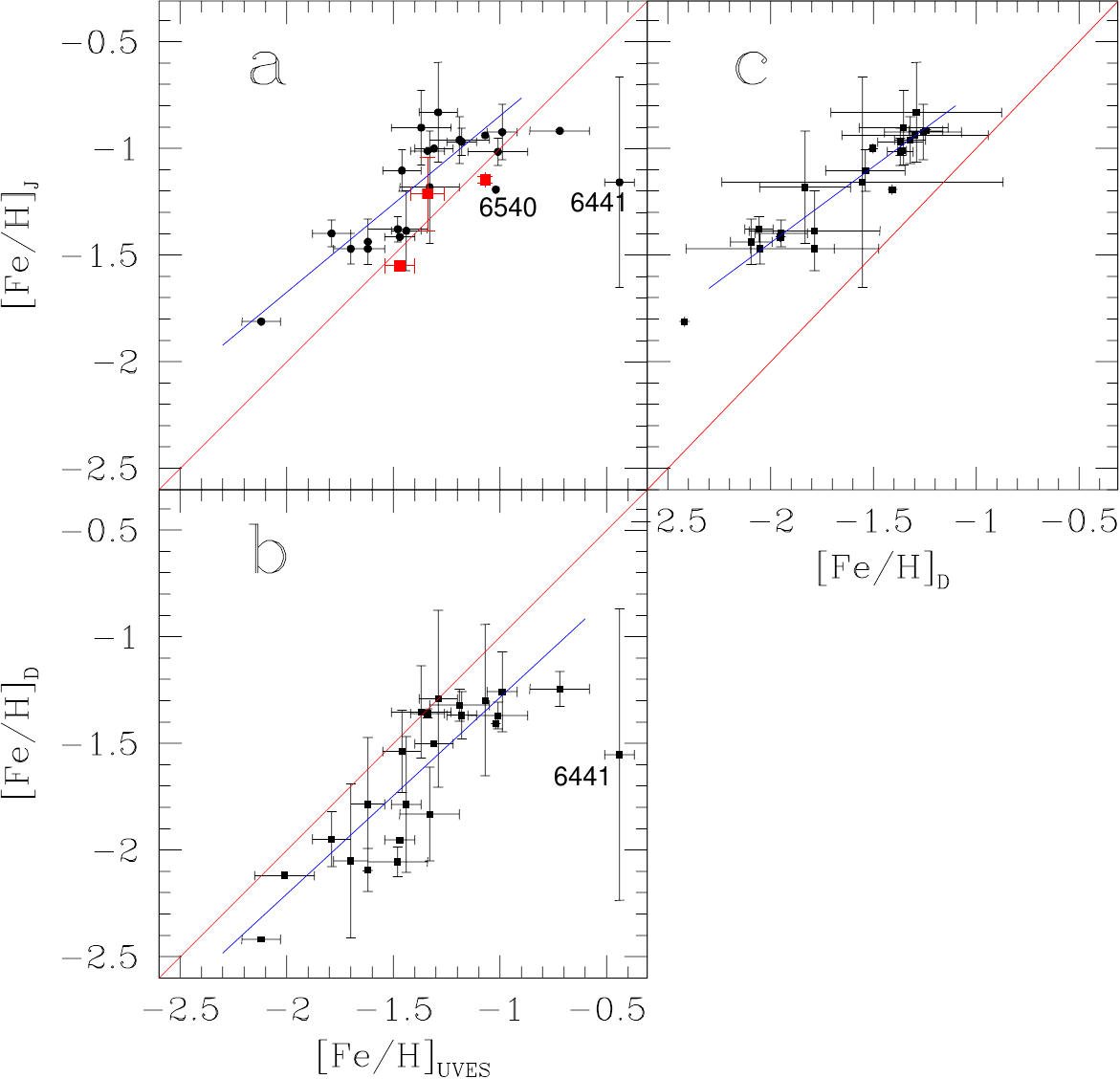}
\caption{Comparison of metallicities. Panel $a)$ [Fe/H]$_J$ vs [Fe/H]$_{\rm UVES}$, panel $b)$ [Fe/H]$_D$ vs [Fe/H]$_{\rm UVES}$ and panel $c)$ [Fe/H]$_J$ vs [Fe/H]$_D$. The blue lines are least-squares fits to the distribution of black points. The systematic shift of [Fe/H]$_J$ and [Fe/H]$_D$ relative to the spectroscopic values [Fe/H]$_{\rm UVES}$ are evident. The red squares in panel $a)$ correspond to the metallicities [Fe/H]$^{\rm UV}_{\rm JK}$ vs [Fe/H]$_{\rm UVES}$ calculated via the Fourier decompositions of OGLE $V$ light curves of RR Lyrae stars satisfying the inclusion criteria in clusters NGC 6522, NGC 6544 and Djorg 2 (see Table \ref{Results2}). They show a better agreement with the spectroscopic results. See \S \ref{sec:Results} for a detailed discussion.}
\label{Comp_Fe}
\end{center}
\end{figure*}

A graphical comparison of our results with these well-established distance, and spectroscopic metalicity scales is performed in Figures \ref{Comp_Distance} and \ref{Comp_Fe}, respectively.

In Fig. \ref{Comp_Distance} we compare the distances $d_I$ and $d_P$ with the critical average distances of \citet{Baumgardt2021} $d_{BV}$. Panels {\it a, b} an {\it c}, show $d_{BV}$ vs. $d_I$, $d_{BV}$ vs. $d_P$ and $d_P$ vs. $d_I$ respectively. In all cases, the agreement is satisfactory. Small average distances are $d_I$-$d_{VB}$=+0.42 kpc, $d_P$-$d_{VB}$=+0.30 kpc and  $d_I$- $d_P$=+0.12 kpc.

In Fig. \ref{Comp_Fe}, a similar comparison is performed for our iron abundance calculations. Panels {\it a, b} an {\it c}, show [Fe/H]$_J$ vs [Fe/H]$_{\rm UVES}$, [Fe/H]$_D$ vs [Fe/H]$_{\rm UVES}$ and [Fe/H]$_J$ vs [Fe/H]$_D$ respectively. Linear fits to the data distribution are shown with blue lines. The outstanding result is that the iron abundances from the calibration of \citet{Jurcsik2021} (eq. \ref{eq:J21}), [Fe/H]$_J$, are larger than the iron values in the spectroscopic scale of \citet{Carretta2009} by an average of 0.145 dex. Likewise the
iron abundances from the calibration of \citet{Dekany2021} (eqs. \ref{eq:D1} and \ref{eq:D2}), [Fe/H]$_D$, are systematically smaller than the spectroscopic scale of \citet{Carretta2009} by an average of -0.297 dex. A similar difference in photometric metallicities derived from \citet{Smolec2005} and \citet{Dekany2021} has also been noted in \citet{Dekany2021}. This is likely due to a different metallicity scale where \citet{Dekany2021} calibration relies on high-resolution spectroscopic metallicities of field RR~Lyrae stars estimated in \citet{Crestani2021a,Crestani2021b}.


 We recall that to calculate $d_I$ the metallicity [Fe/H]$_D$ was employed. Since this metallicity scale turned out to be $-0.297$ dex poorer relative to the UVES spectroscopic scale, the question is how this is affecting the calculation of the distance. One can corroborate that a metallicity difference $-0.297$ in eq.~5 produces a brighter $M_I$ of about $0.06$ mag and hence a larger distance by about $230$\,pc \citep[at the distance to the Galactic bulge, $R_{0}=8178$\,pc][]{GRAVITY2019}. Hence, the distance differences $d_I$-$d_{VB}=+0.42$\,kpc and $d_P$-$d_{VB}=+0.30$\,kpc are nearly completely attributable to the drift in metallicity [Fe/H]$_D$. 

The overabundance of the photometric metallicities [Fe/H]$_J$ has already being pointed out by \citet{Jurcsik2021}: "The application of this transformation [calibrations] does not solve the problem of the overestimation of the photometric [Fe/H] for most metal-poor GC,...".

We point out that the calibrations for RRab and RRc stars of \citet{Jurcsik1996} and \citet{Morgan2007}, respectively, for the Fourier decomposition of $V$-band light curves, when transformed to the UVES scale (using the transformation from \citet{Jurcsik1995}), produce iron abundances in good agreement with the spectroscopic scale of \citep{Carretta2009}, as has been shown by \citet{Arellano2024} (his Figs. 1a and 1b) for a family of 39 globular clusters.

In order to further revise these statements, we recover OGLE high-quality $V$-light curves, with more than 120 measurements and well covered in phase, for stars that satisfy the four criteria defined in \S \ref{inclusion}, in three clusters; NGC 6522 (V1, V3, V5); NGC 6544 (V2) and Djorg 2 (V2, V10). These light curves were Fourier decomposed and then proceeded as described in \S 3 of \citet{Arellano2024}; the calibration for [Fe/H]$_{JK}$
for RRab stars \citep{Jurcsik1996} and for [Fe/H]$_{ZW}$ for RRc stars \citep{Morgan2007} (eqs. 2 and 3) were employed and the metallicities were transformed to the UVES scale of \citet{Carretta2009}. We shall refer to these metallicities as [Fe/H]$^{\rm UV}_{\rm JK}$. For the distance we used the $M_V$ calibrations of \citet{Kovacs2001} and \citet{Kovacs1998} for RRab and RRc stars respectively (eqs. 6 and 7 in \citet{Arellano2024}. The results are given in Table \ref{Results2}.
We over plotted the [Fe/H]$^{\rm UV}_{\rm JK}$ vs [Fe/H]$_{\rm UVES}$  in panel $a$ of Fig. \ref{Comp_Fe} with red squares. It is clear that the agreement with the spectroscopic scale [Fe/H]$_{\rm UVES}$ is more satisfactory.

In conclusion, the iron calibrations from $I-band$ light curves decompositions of eqs. \ref{eq:J21} \citep{Jurcsik2021}, \ref{eq:D1} and \ref{eq:D2} \citep{Dekany2021} produce iron values richer by an average of 0.145 dex and poorer by -0.297 dex, respectively, than the spectroscopic scale of \citet{Carretta2009}. On the other hand, we have confirmed that the iron calibrations for RRab stars \citet{Jurcsik1996} and for RRc stars \citet{Morgan2007} and those for the absolute magnitudes of \citet{Kovacs2001} and \citet{Kovacs1998} for RRab and RRc stars respectively,
produce metallicities and distances in agreement with the highly  respected spectroscopic UVES scale \citep{Carretta2009} and distances \citep{Baumgardt2021}, and may not need a deep revision.

\begin{table*}
\centering
    \caption{Metallicity and distance from OGLE $V$-band light curves}
\begin{tabular}{ccccccc}
\hline
Cluster & [Fe/H]$^{UV}_{JK}$  & d &Nab & Nc & [Fe/H]$_{\rm UVES}$ & d$_{BV}$ \\
NGC &  & kpc& & & & kpc   \\
\hline

6522&  -1.213$\pm$0.173 &9.19$\pm$0.35 & 0 & 3 &-1.34$\pm$0.08  &7.30$\pm$0.21\\

6544&-1.55   & 2.92   &1 &0   & -1.47$\pm$0.07  &2.58$\pm$0.60 \\

Djorg 2& -1.147$\pm$0.017 &7.71$\pm$0.39  &0 & 3  & -1.07+ &8.76$\pm$0.18 \\
\hline

\end{tabular}

\label{Results2}
\center{+ Value from \citet{Schiavon2024}} 

\end{table*}

\vskip 1.0cm

\section{Bulge cluster spatial and kinematical distribution}

Following the approach adopted in \citet{Prudil2025Kin}, we computed orbits for our sample of bulge globular clusters by evolving their phase-space coordinates in an analytical, non-axisymmetric Milky Way potential. As input, we adopted our newly derived distance estimates (using $d_P$ from Table~\ref{Tab:DistanceToClusters}), together with proper motions, sky coordinates, and line-of-sight velocities taken from the Galactic globular cluster database\footnote{Provided here \url{https://people.smp.uq.edu.au/HolgerBaumgardt/globular/}.}, following \citet{Vasiliev2021,Baumgardt2021}.

The Galactic potential corresponds to the barred Milky Way model of \citet{Portail2017Pattern}, as implemented and extended by \citet{Sormani2022} and \citet{Hunter2024} within the \texttt{AGAMA} framework\footnote{Available here \url{https://github.com/GalacticDynamics-Oxford/Agama}.} \citep{Vasiliev2019Agama}. This model includes a rotating bar and axisymmetric components representing the nuclear regions, stellar disks, and dark matter halo, and was evolved with a fixed pattern speed, as described in \citet{Prudil2025Kin}. Each orbit was integrated for a total time of $5\,\mathrm{Gyr}$, chosen to ensure robust characterization of orbital properties. 

\begin{figure}
\centering

\includegraphics[width=\columnwidth]{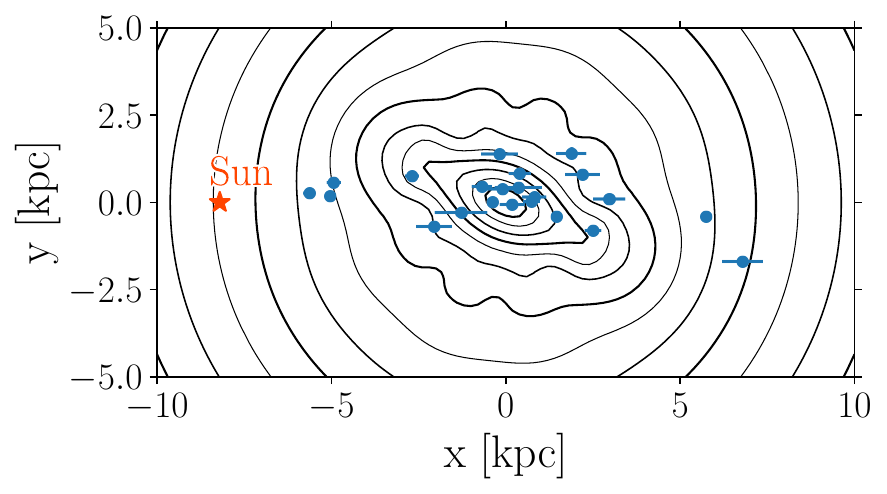}
\includegraphics[width=\columnwidth]{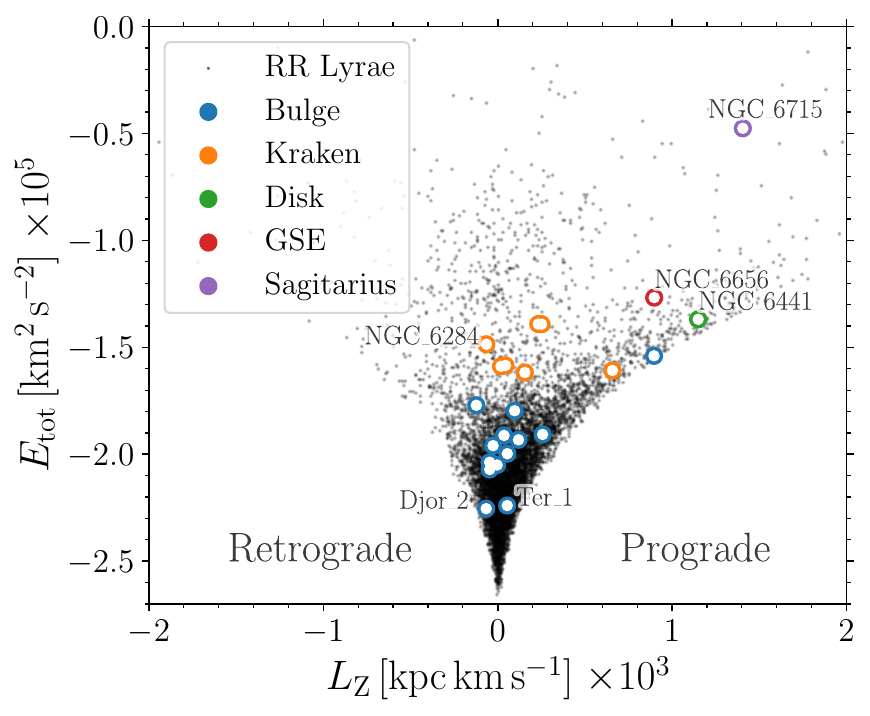}
\caption{The spatial distribution of the analyzed globular clusters in Cartesian coordinates, using mean distances from \citet{Prudil2025} in the top panel, and orbital energy vs. angular momentum in the bottom panel. The black lines in the top panel show surface-density contours of an analytical potential implemented in the \texttt{AGAMA} software \citep{Vasiliev2019Agama}, for a stellar component rotated by a bar angle of 25$^\circ$. The colored points in the bottom panel depict analysed globular clusters \citep[black points representing RR~Lyrae stars from][study]{Prudil2025Kin} in this study, with color-coding based on association with one of the MW substructures or accreted structures. For a handful of displayed globular clusters, we provide their associated labels.}
\label{fig:ClustersBar}
\end{figure}

We used the $d_P$ distances listed in Table~\ref{Tab:DistanceToClusters} to illustrate the Galactic distribution of our bulge globular cluster sample in the Cartesian plane (top panel of Fig.~\ref{fig:ClustersBar}). For consistency with the analysis presented in Sect.~\ref{sec:Results}, these distances correspond to the mean $d_P$ values computed from \citet{Prudil2025} using only stars that satisfy the four selection criteria defined above (with the corresponding number of stars given in parentheses in the table). The clusters are predominantly located within the central few kiloparsecs of the Galaxy, providing a spatial view of their distribution in the inner Milky Way.

In the bottom panel of Fig.~\ref{fig:ClustersBar}, we show the distribution of orbital energy as a function of angular momentum for the analysed globular clusters. Using the association analysis of \citet{Callingham2022}, we highlight the potential origins of our globular cluster sample. The color-coded distribution in Fig.\ref{fig:ClustersBar} provides additional insight into the likely origin of the globular cluster population studies in this work. Clusters associated with the in situ bulge component (blue symbols) occupy a relatively narrow locus at low angular momentum and tightly clustered binding energies, consistent with formation in the early inner Milky Way and subsequent evolution dominated by the barred potential. In contrast, clusters linked to accreted progenitors span a broader region in the $E_{\rm tot}$ vs. $L_{\rm Z}$ plane, including both prograde and retrograde orbits, reflecting the diverse orbital properties expected for systems deposited by past merger events. This behavior closely mirrors the trends identified in the integrals-of-motion analysis of Galactic globular clusters and field RR~Lyrae stars, where accreted populations exhibit a much larger spread in orbital energy and angular momentum than their in situ counterparts \citep[e.g.,][]{Callingham2022,Luongo2024}. 

The small differences, most notably on the prograde side at $E_{\rm tot} \approx -1.5 \times 10^{5}$, between the orbital properties presented here and those reported in other studies \citep[e.g., using axisymmetric potential,][]{Callingham2022,Luongo2024} can be attributed to differences in the adopted Galactic potential for orbital integration, as well as to variations in the assumed distances (e.g. for the clusters NGC~6540 and NGC~6441).



\section{Summary and Conclusions}

The metallicities of cluster member RR Lyrae stars, obtained from their light curve Fourier decomposition, were calculated using recent calibrations for the $I$-band \citep{Jurcsik2021, Dekany2021}. We used for this purpose the $I$-light curves from the OGLE IV database. We found that the [Fe/H] values of the calibrations of \citet{Jurcsik2021} and \citet{Dekany2021} are on average $+0.145$ and $-0.297$ dex, respectively, compared to the spectroscopic values on the UVES scale \citep{Carretta2009}. We remark that the photometric calibrations using the $V$-band Fourier decompositions \citep{Jurcsik1996,Morgan2007} for RRab and RRc stars, respectively, produce metallicities in good agreement with the UVES scale \citep{Arellano2024}.

The MPZ distances, $d_P$, averaged exclusively for cluster member stars, agree within 0.45 kpc with the cluster distance catalogue of \citet{Vasiliev2021}.

Employing the PMZ distances for member RR~Lyrae stars in each cluster, we investigated the spatial distribution and orbital properties of the bulge globular cluster sample by integrating their orbits in an analytical, non-axisymmetric Milky Way potential that includes a rotating bar. The clusters are predominantly located within the central few kiloparsecs of the Galaxy, providing a clear spatial view of their distribution in the inner Milky Way.

In the orbital energy versus angular momentum plane, the globular clusters separate into distinct groups consistent with their proposed origins: clusters associated with the in situ bulge population occupy a narrow locus at low angular momentum and tightly clustered binding energies, while clusters linked to accreted progenitors span a much broader range of orbital energies and include both prograde and retrograde orbits. These trends closely mirror those found in previous integrals-of-motion studies of Galactic globular clusters and field RR~Lyrae stars, with the small differences observed attributable to the adopted barred Galactic potential and to variations in the assumed cluster distances.

\section*{Acknowledgments}

The present project has been partially supported by DGAPA-UNAM through project  IN103024. AAF is grateful to the European Southern Observatory (ESO) for hospitality in Garching in 2025.

\section*{DATA AVAILABILITY}
The data used in this work are available in the OGLE IV database. The results in the form of tables can be shared on request to the corresponding
author.

\bibliographystyle{mnras}
\bibliography{BULGE}

\appendix
\section{Discussion on individual clusters}

\label{sec:appendix}

{\bf NGC 6266}. The stars V37 and V94 are nearly 2 mag brighter than average HB. The stars are not a cluster members in spite of m1 or m2 code.

V227 is about 1.6 mag fainter than average HB. Not a member in spite of its m3 code. 

{\bf NGC 6293}.  Only RRc's appear to be a cluster member (V6).

{\bf NGC 6355}. It has four member stars. We employed the four of them for the calculations. We should note that the distance of V1 given by \citet{Prudil2025} is anomalously small (6727.83), in spite of having <$I_o$> similar to the other three. 
Its anomalous distance may be due to the fact that it was based on $K$ and $J$ bands, but the other three stars have distances based on OGLE and Gaia.
This lowers the average $d_P$.  In the calculation using the average reddening from Harris it gives a good distance consistent with the other member stars.

{\bf NGC 6401}. Star V4 (OGLE-BLG-RRLYR-24049) is another one of those havig anomalous $d_P$ but regular $I_0$, and $d_I$. Two more RRc stars; V28 (OGLE-BLG-RRLYR-23999) and V30 (OGLE-BLG-RRLYR-24021) display a large scatter in the $I$ light curve. They may be double mode stars. It is worth exploring that possibility. They were not included in the calculations for Table 1.

{\bf NGC 6441}. Among the RRLs labeled as cluster member in \citet{Clement2001}, V65 (OGLE-BLG-RRLYR-03957) is 2 mag below the average HB and hence it is not a cluster member. Some stars in the HB have anomalous distances in \citet{Prudil2025}; V64 (OGLE-BLG-RRLYR-03912), V124 (OGLE-BLG-RRLYR-03913), or have no distance estimation. These were not considered in the average calculation $d_P$. For the Fourier calculations those stars exhibiting a very large scatter in the $I$ light curves were not included, these are V96 (OGLE-BLG-RRLYR-30154), V115 (OGLE-BLG-RRLYR-03962), V123 (OGLE-BLG-RRLYR-30200) and V124 (OGLE-BLG-RRLYR-03913).

{\bf NGC 6453}. It contains two RRab (V4, V8) and one RRc (V9). The latter displays a large scatter and we suspect it is a double mode star. Hence it was eliminated from the calculation of $d_P$.  The three $d_I$ distances are consistent.

V9 (OGLE-BLG-RRLYR-30442), has ipd\_frac\_multi\_peak equal to 10 - it might be blended, RUWE flag from Gaia is also above 1.4. The difference in apparent magnitude in the $K$-band between this star and the other two member stars is only 0.2 mag. Also, its photometric metallicity $\sim$ -1.5 differs from the metallicity of the other two $\sim$ -2.0 dex. The stars does not seem to have additional potential modes \citep{Netzel2019}. The bad photometry may come from its proximity to the cluster-dense environment. The star was not included in the averages but only the other two members V4 and V8.

{\bf NGC 6401}. V4, V28 and V30 have odd distances. In spite of their proper motions suggesting their membership, they are off the HB. They were not considered in the calculations of $d_I$ and $d_P$. However, inspite their large "Blazhko and/or scatter the $d_I$
distances are reasonable and not so peculiar as the $d_P$'s. We opted for not considering them in either average.

{\bf NGC 6522}. The standar distance is about 8.5 kpc. V4 (OGLE-BLG-RRLYR-12132), V10 (OGLE-BLG-RRLYR-12115) and V11 (OGLE-BLG-RRLYR-12114) give odd distances, both $d_I$ and $d_P$. V4 and V10 are clearly off the HB hence they are not authentic members. For V11  there is no OGLE photometry. They were not considered in the average calcualtion of $d_I$ and $d_P$.

{\bf NGC 6540}. There is only one cluster member RRab star, V2.
The distance for this star from \citet{Prudil2025} database is a bit short, 3.17 kpc compared with the result from the Fourier decomposition (5.83 kpc) and the one adopted by Baumgard et al. (5.91±0.27 kpc). It was found that the star is blended and has $K$-band apparent magnitude of 12.6 produces too short a distance.  Its $Gaia$ $I$-band distance is around 8 kpc. While the star is peculiar we
decided to keep it for comparison purposes.

{\bf NGC 6544}. It contains only one RRab which is likely a cluster member. 

{\bf NGC 6558}. Four member stars all in the HB without peculiarities.

{\bf NGC 6569}. There are 13 cluster members \citet{Clement2001}, all within 0.2 mag of the average HB, which confirms their membership. However, V18 (OGLE-BLG-RRLYR-34973), V27 (OGLE-BLG-RRLYR-34970), V28 (OGLE-BLG-RRLYR-34976), V30 (OGLE-BLG-RRLYR-34978), and V32 (OGLE-BLG-RRLYR-34993), have odd distances in \citet{Prudil2025} and were not included in the average $d_P$, which otherwise matches well the value of $d_I$ and $d_{BV}$.

{\bf NGC 6626}. There are 11 RR Lyrae labeled as cluster members in \citet{Clement2001}. V20 (OGLE-BLG-RRLYR-68201) is nearly 2 mag off the average HB, hence it is not an authentic cluster member. V26 (OGLE-BLG-RRLYR-59825) has no distance estimate in \citet{Prudil2025}. 

{\bf NGC 6638}. This cluster also has one star (V30: OGLE-BLG-RRLYR-62064) with too short $d_P$ in spite of having regular brightness level and a reasonable $d_I$. However, its light curve is scarce, hence it was also not include in the Fourier [Fe/H] and $d_I$ calculations.

{\bf NGC 6642}. The star V16 (OGLE-BLG-RRLYR-62367) displays a very strong Blazhko modulation. It was not included in our calculations.

{\bf NGC 6656}. Only V6 (OGLE-BLG-RRLYR-36669) was excluded from the $d_I$ calculation due to large Blazhko modulations. 

{\bf NGC 6681}. There are three member stars. However, one (V6: OGLE-BLG-RRLYR-65596)  has a very scattered light curve, hence it was not included in the calculations.

{\bf NGC 6715}. The mean $I$ level of the HB is $\sim$17.5 mag. V192, V240, V244 and  V246 outstand for being 1 to 2.5 mag. above the HB, hence they are not authentice cluster members and were ignored for the calculations. The remaining 91 RR Lyrae stars, labeled as cluster members in the CVSGC, are within 0.2-0.3 mag of the HB. However, some of these exhibited very strong Blazhko-like amplitude modulations and were not included. In the end, 61 RRab and 11 RRc were included in the Fourier decomposition calculations shown in Table 1. Among the 91 member RR Lyrae stars, some have either no distance estimation or a spurious value in the database of \citet{Prudil2025}, which were not included in the average $d_P$ (V4, V31, V33, V40, V46, V51, V67, V78 , V80, V82, V86, V88, V97, V124, V125, V127, V129, V139, V140, V160, V162, V163, V172, V176, V180,  V188, V240, V244, V246, V250, V252, V295, V308). Most of them, 25 out of 33, do not have enough mean magnitudes in different passbands. The remaining have distances, but some are blended. It should be noticed that many of these stars display regular OGLE $I$ light curves that lead to reasonable distances.
 
{\bf Terzan 1}. Stars V8, V11, V13 and V16, with no distance estimation or spurious value in \citet{Prudil2025} were not included in the $d_P$ average. The nine member stars were included for the $d_I$ calculation.

{\bf Djorg 2}. It contains one RRab and one RRc stars, both in the HB and labeled as cluster members. Their distances in \citet{Prudil2025} are discrepant: 9.22 and 7.22 kpc respectively. Their $I$ light curve Fourier decomposition lead to distances 8.26 and 8.42 kpc respectively.

\bsp	
\label{lastpage}
\end{document}